\documentclass[11pt]{extarticle}
\usepackage{fullpage}
\usepackage[english]{babel}
\usepackage[utf8]{inputenc}
\usepackage{amsmath, amssymb, bm}
\usepackage{graphicx}
\usepackage{subcaption}
\usepackage[colorinlistoftodos]{todonotes}
\usepackage{indentfirst}

\newcommand{\hiddenpower}[2] { \ifnum \numexpr#2=1 #1 \else #1^#2 \fi }
\numberwithin{equation}{section}

\usepackage{xparse}
\ExplSyntaxOn
\newcounter{diff_order}
\newcounter{diff_power}

\newcommand{\rawdiff}[3]
{
	\setcounter{diff_order}{0}
	\clist_map_inline:nn{#3}{\stepcounter{diff_order}}
	
	\frac{\hiddenpower{#1}{\thediff_order} #2}
	{
		\def\old_var{DefaultValue}
		\setcounter{diff_power}{0}
		
		\clist_map_inline:nn{#3}
		{
			\def\new_var{##1}
			\ifnum \thediff_power=0
				\stepcounter{diff_power}
			\else
				\tl_if_eq:NNTF \new_var \old_var
				{\stepcounter{diff_power}}
				{
					#1 \hiddenpower{\old_var}{\thediff_power}
					\setcounter{diff_power}{1}
				}
			\fi

			\def\old_var{##1}
		}
		
		#1 \hiddenpower{\old_var}{\thediff_power}
	}
}

\ExplSyntaxOff

\newcommand{\lb}{\left(}
\newcommand{\rb}{\right)}

\renewcommand{\ln}[1]{\text{ln} \lb #1 \rb}

\begin{document}


\begin{center}
\strut\hfill



\noindent {\LARGE{\bf{Time-like boundary conditions in the NLS model}}}\\
\vskip 0.3in

\noindent {\bf {{Anastasia Doikou, Iain Findlay and Spyridoula Sklaveniti}}}
\vskip 0.4in

\noindent {\footnotesize School of Mathematical and Computer Sciences,
Department of Mathematics,\\ Heriot-Watt University,
Edinburgh EH14 4AS, United Kingdom}

\vskip 0.1in
\noindent {\footnotesize {\tt E-mail: a.doikou@hw.ac.uk,\ iaf1@hw.ac.uk,\ ss153@hw.ac.uk }}\\

\vskip 0.60in

\end{center}

\begin{abstract}
\noindent 
We focus on the non-linear Schr\"odinger model  and we extend the notion of space-time dualities in the presence of integrable time-like boundary conditions. We identify the associated time-like ``conserved'' quantities 
and Lax pairs as well as the corresponding boundary conditions. In particular, we derive the generating function of the space components of the Lax pairs in the case of 
time-like boundaries defined by solutions of the reflection equation. Analytical conditions on the boundary Lax pair lead to the time like-boundary conditions. 
The time-like dressing is also performed for the first time, as an effective means to
produce the space components of the Lax pair of the associated hierarchy.  This is particularly relevant in the absence of a classical  $r$-matrix, 
or when considering complicated underlying algebraic structures. The associated time Riccati equations and hence the time-like conserved quantities are also derived. We use as the main paradigm for this purpose the matrix NLS-type hierarchy.
\end{abstract}

\date{}
\vskip 0.4in



\section{Introduction}

\noindent 
The non-linear Schr\"odinger (NLS) model is one of the most well studied integrable models at the classical and quantum level 
(see e.g \cite{Manakov}--\cite{ADP}). A considerable amount of work is devoted to the study of the model
from the algebraic/Hamiltonian standpoint \cite{FT}  in the case of periodic as well as generic integrable boundary conditions \cite{Sklyanin, DoFioRa}. 
By means of the algebraic formulation one can systematically construct the infinite tower of conserved quantities as well as
the hierarchy of Lax pairs via the generating function of the time components of the Lax pairs based on the existence of a classical $r$-matrix \cite{STS}. 
The universal formula that provides the generating function of the time components of the Lax pairs was also derived in  the case of generic 
integrable boundary conditions in \cite{AvanDoikou_boundary, DoikouFindlay}.

Thus far only space-like integrable boundary conditions have been considered, whereas the issue of time-like boundaries has not been systematically addressed.
We focus here on the NLS model and we generalize the idea of space-time dualities studied in 
\cite{CaudrelierKundu, ACDK} in the case of generic time-like integrable boundary conditions. To achieve this we implement the Hamiltonian description based on the existence 
of the classical $r$-matrix, providing the underlying Poisson  structure. As argued in \cite{ACDK} the time-like Poisson structure gives rise to an ultra-local Poisson algebra for
the $t$ component of the Lax pair, with the same classical $r$-matrix as the ultra-local Poisson algebra satisfied by the $x$ component with respect to the usual Poisson bracket. 
We first briefly review the results found in \cite{ACDK} for integrable dual periodic systems by introducing time-like Poisson algebras.
We then  move on to the case of generic integrable boundary conditions, and based on the fundamental algebraic relations ($t$-Poisson)
we extend the idea of Sklyanin's modified monodromy \cite{Sklyanin} along the time axis.  We produce novel results regarding time-like integrable boundary conditions via the
derivation of the respective ``conserved'' quantities and the $x$-part of the Lax pairs in the presence of general open boundaries. 

We also derive for the first time the time like dressing
process and we produce the hierarchy of conserved quantities as well as the space components of the Lax pairs of the matrix AKNS (NLS-type) hierarchy. 
In this frame we also identify the time Riccati  equation associated to the generic time-like dressing transform, which is equivalent to the Riccati equation for the solution of the time part of the auxiliary linear problem. The relevant time-like conserved quantities are also derived.

\section{Time-like integrable boundary conditions}



\noindent 
We focus on the study of space-time dualities and more precisely on the implementation of time-like integrable  boundary conditions, 
extending the results of \cite{ACDK}. Various studies address the issue of integrable  
boundary conditions for the NLS model and its generalizations, 
but in the majority of these investigations space-like boundary conditions are considered. Here we are going to reverse the picture and consider 
time-like boundary conditions exploiting recent results on the time-space duality in the NLS case.
In \cite{CaudrelierKundu, ACDK}  the concept of ``dual" integrable 1+1 dimensional models was introduced, specifically in reference to the NLS model. 
We focus our attention here on the time-like version of the NLS hierarchy and extend the description of \cite{ACDK} in the presence of integrable 
time-like boundary conditions. 

Let us first recall the space-like description. The starting point is the $U$-operator of the Lax pair $\big (U,\ V\big) $ consisting of generic $c$-number 
$d \times d$  matrices (see e.g. \cite{FT}). The Lax pair matrices depend in general on some fields and a spectral parameter, and obey the auxiliary linear problem:
\begin{eqnarray}
	&&\partial_x \Psi(\lambda, x,t) = U(\lambda, x,t) \Psi(x,t),  \nonumber\\
 && \partial_{t} \Psi(\lambda, x,t)= V(\lambda, x,t) \Psi(\lambda, x,t). \label{ALP}
\end{eqnarray}

For the NLS-type system in particular the $U,\  V$ matrices are given as
\begin{equation}
	U(\lambda, x,t) = \lb \begin{matrix}
		\frac{ \lambda}{2}& \hat   u \\
		 u & -\frac{ \lambda}{2}
	\end{matrix} \rb, ~~~~V(\lambda, x, t) = \lb \begin{matrix}
		\frac{ \lambda^2}{2}- u \hat u& \lambda \hat   u + \partial_x \hat u\\
		 \lambda u - \partial_x u & -\frac{ \lambda^2}{2} + u \hat u
	\end{matrix} \rb, \label{eq:NLS_Lax}
\end{equation}
where the fields $u,\ \hat u$ depend on $x,\  t$.

Assume that the $U$-operator satisfies the linear Poisson structure 
\begin{equation}
	\Big\lbrace U_1(x, \lambda),\  U_2(y, \mu) \Big\rbrace_{S} = \Big[ r_{12}(\lambda - \mu),\  U_1(x, \lambda) + U_2(y, \mu) \Big] \delta(x - y), \label{eq:LinAlg}
\end{equation}
where the $r$-matrix is a solution of the classical Yang-Baxter equation \cite{STS}, and the subscript $_S$ denotes space-like Poisson structure.  Equation (\ref{eq:LinAlg}) acts on 
${\mathbb V} \otimes {\mathbb V}$, where ${\mathbb V}$ is in general a $d$ dimensional space, and the indices $1,\ 2$ in 
(\ref{eq:LinAlg}) denote the first and second space respectively. In general, for any $d \times d$ matrix $A$ the quantities $A_1,\  A_2$ are defined 
as $A_1 = A\otimes {\mathbb I},\ A_2= {\mathbb I} \otimes A$ i.e. $ A_1$ acts non-trivially on the first space, whereas $A_2$ acts on the second one, with
${\mathbb I}$ the $d\times d$  identity matrix. 
The $r$-matrix acts on both spaces, and for the particular example we are going to examine here, $r$ is the Yangian solution \cite{Yangian},
\begin{equation}
r_{12}(\lambda) ={1\over\lambda}\ \sum_{i, j=1}^{d} e_{ij} \otimes e_{ji}, \label{YangS}
\end{equation}
where $e_{ij}$ are $d \times d$ matrices with elements $(e_{ij})_{kl} = \delta_{ik} \delta_{jl}$. The quantity $\sum_{i, j} e_{ij} \otimes e_{ji}$  
is the so called permutation operator. Then recalling (\ref{eq:NLS_Lax}),  (\ref{eq:LinAlg})  and (\ref{YangS}) we conclude that 
\begin{equation}
\Big \lbrace u(x),\ \hat u(y)\Big \rbrace_S=\delta(x-y).
\end{equation}
It is worth noting that in the space-like formulation the $U$-matrix (\ref{eq:NLS_Lax}) is the starting point and the conserved 
quantities as well as the hierarchy of $V$-operators emerge from it \cite{STS,  DoFioRa}. In the time-like approach 
on the other hand the starting point is some $V$-operator, and from this the time-like conserved quantities as well as the 
$U$-hierarchy are derived \cite{ACDK}.

The key object in this setting is the space monodromy, a solution of the first of the equations (\ref{ALP}),
\begin{equation}
T_{S}(a, b, \lambda) = P\exp\Big ({\int_{b}^{a} U(x, \lambda) dx}\Big ),  ~~~~a > b,\nonumber \label{xmono}
\end{equation}
which satisfies a quadratic algebra, and guarantees space Poisson commutativity and thus integrability:
\begin{equation}
\Big \{trT_S(\lambda),\ trT_S(\mu) \Big \} _S=0.
\end{equation}

In \cite{CaudrelierKundu, ACDK} the picture was reversed, that is,  it was assumed that $V$, as well as $U$, satisfies a linear algebra 
(see also \cite{AvanCaudrelier} on further emphasis on the algebraic/$r$-matrix description).
Indeed, it was noticed  in \cite{ACDK} that the time-like Poisson bracket could be constructed 
from an equivalent linear algebraic expression regarding the time component of the Lax pair:
\begin{equation}
	\Big\lbrace V_1(t_1, \lambda),\  V_2(t_2, \mu) \Big\rbrace_T= \Big[ r_{12}(\lambda - \mu),\ 
V_1(t_1, \lambda) + V_2(t_2, \mu) \Big] \delta(t_1 - t_2), \label{Dual2}
\end{equation}
where $r$ is the same classical $r$-matrix as in (\ref{eq:LinAlg}), and the subscript $_T$ denotes the time-like Poisson structure. 
Then the time monodromy $T$, a solution to the time part of \eqref{ALP} is
\begin{equation}
T_T(a, b, \lambda) = P\exp\Big ({\int_{b}^{a} V(t, \lambda) d t}\Big ),  ~~~~a > b \label{tmono}
\end{equation}
and satisfies the quadratic algebra
\begin{equation}
\Big \{ T_{T1}(\lambda),\ T_{T2}(\mu) \Big \}_T= \Big [ r_{12}(\lambda -\mu),\ T_{T1}(\lambda) T_{T2}(\mu) \Big ]. \label{quad}
\end{equation}
Consequently one obtains commuting operators, with respect to the time-like Poisson structure
\begin{equation}
\Big \{trT_T(\lambda),\ trT_T(\mu) \Big \}_T=0.
\end{equation}
\noindent

Inspired by the form of the $V$-operator for the NLS model we express our starting operator $V$ in the following form ($d=2$):
\begin{equation}
 V(\lambda) = \lb \begin{matrix}
		\frac{ \lambda^2}{2}  -  u\hat  u & \lambda \hat u + \pi\\
		\lambda  u - \hat \pi & - \frac{ \lambda^2}{2}+  u\hat  u
	\end{matrix} \rb. \label{NLSb}
\end{equation}
We  require $V$ to satisfy the time-like Poisson structure  (\ref{Dual2}) and we then produce the time-like algebra for the fields,  
which reads as (we only write below the non zero commutators, see also \cite{ACDK}):
\begin{eqnarray}
  \Big \lbrace u(t),\ \pi(t') \Big \rbrace_T  = \Big \lbrace \hat  u(t),\ \hat \pi(t') \Big \rbrace_T  = \delta(t-t').
\end{eqnarray}
Henceforth,  we focus only on time-like Poisson structures thus we drop the subscript $_T$ whenever this applies.

\subsection{Periodic Boundary Conditions}
\label{sec:Periodic}

\noindent 
We start by briefly deriving the results found in \cite{ACDK} for dual systems with periodic boundary conditions, in the language of Lax pairs. 
The starting point for this construction is the auxiliary linear problem, \eqref{ALP}, and the algebraic relation \eqref{Dual2}. 
Here we exclusively discuss time-like boundary conditions,  however note that space-like boundaries have been 
discussed from the Hamiltonian point of view in \cite{Sklyanin, DoFioRa}.

The key object in our analysis is the time monodromy matrix (\ref{tmono}), which can be decomposed as
\begin{equation}
T(t,t';\lambda) = \big (1+ {\mathrm W}(t; \lambda)\big )\  e^{{\mathrm Z}(t, t'; \lambda)}\ \big (1+{\mathrm W}(t'; \lambda)\big)^{-1}, ~~~t>t', \label{deco}
\end{equation}
where ${\mathrm W}$ is anti-diagonal and ${\mathrm Z}$ is purely diagonal. Then one obtains the typical Riccati equation for ${\mathrm W}$
\begin{eqnarray}
&& \partial_t {\mathrm W} + \Big [ {\mathrm W},\  V_D\Big ] + {\mathrm W}V_A {\mathrm W} - V_A =0, \label{tRiccati}\\
&& \partial_t {\mathrm Z} =V_D +V_A {\mathrm W}, \label{riccati1}
\end{eqnarray}
where $V_D,\ V_A$ are the diagonal and anti-diagonal parts of the operator $V$.
Solutions of the pair of equations above are given in the Appendix for the first several members of the $\lambda$ expansion, i.e. after considering: 
${\mathrm W} = \sum_n {{\mathrm W}^{(n)} \over \lambda^n},\  {\mathrm Z} = \sum_n {{\mathrm Z}^{(n)} \over \lambda^n}$. It is worth writing the Riccati equations 
for the element of the matrix ${\mathrm W}$
\begin{eqnarray}
&&\partial_t {\mathrm W}_{21}= \lambda u -\hat \pi +(2 u \hat u - \lambda^2)  {\mathrm W}_{21} -(\pi +\lambda \hat u) {\mathrm W}_{21}^2 \label{rr1}\\
&& \partial_t {\mathrm W}_{12} = \lambda \hat u + \pi +(-2 u \hat u + \lambda^2)  {\mathrm W}_{12}+ (\hat \pi -\lambda  u) {\mathrm W}_{12}^2. \label{rr2}
\end{eqnarray}

Taking the trace and logarithm of the time-like monodromy, we find the generator for an infinite tower of conserved\footnote{Conserved with respect to 
time variations for the space-monodromy matrix, and ``conserved''  with respect to spatial variations for the monodromy matrix built using $ V $.} 
quantities associated to the system:
\begin{equation}
	\mathcal{G}(\lambda) = \ln{tr \big (T (\tau, -\tau, \lambda)\big )}. \nonumber
\end{equation}
Taking into consideration the decomposition of $T$ in (\ref{deco}) as well as the fact that the leading contribution in $e^{\mathrm Z}$ comes from the 
$e^{ {\mathrm Z}_{11}}$ term as $\lambda \to \infty$  (see the expression for ${\mathrm Z}^{(-2)}$ in the Appendix), then we conclude that 
$\mathcal{G}(\lambda) = {\mathrm Z}_{11}(\lambda)$, having also assumed vanishing or periodic boundary conditions at $\pm \tau$.

As in the space-like description we may derive the generating function that provides the hierarchy of $U$-operators associated to each one of the time-like Hamiltonians. 
Indeed, taking into consideration the zero curvature condition as well the time-like Poisson structure satisfied  by $V$ one can show that the generating function of the 
$U$-components of the Lax pairs is given by  (see also \cite{ACDK})
\begin{align}
{\mathbb U}_2(t, \lambda, \mu) &= \mathfrak{t}^{-1}(\lambda) tr_1\Big (T_1(\tau, t, \lambda) r_{12}(\lambda - \mu) T_1(t, -\tau, \lambda)\Big ), \label{eq:UGen}
\end{align}
where $\mathfrak{t}(\lambda) = tr\big(T(\lambda)\big )$. In the case where the $r$-matrix is the 
Yangian (\ref{YangS}), and after taking into consideration the decomposition (\ref{deco}) the latter expression becomes
\begin{eqnarray}
{\mathbb U}(t, \lambda, \mu) &=& {\mathfrak{t}^{-1}(\lambda) \over \lambda -\mu}\  T(t, -\tau, \lambda) \ T(\tau, t, \lambda)  \nonumber\\ 
& =& {1\over \lambda- \mu} \  \big (1+{\mathrm W}(t, \lambda)\big ) D\big  (1+{\mathrm W}(t, \lambda)\big )^{-1}, \label{eq:UGen2}
\end{eqnarray}
where $D= \mbox{diag}(1,\ 0)$.

Now using the expression for the generating function ${\mathcal G}(\lambda) ={\mathrm Z}_{11}(\lambda)$ and the findings presented in the 
Appendix we identify the first couple of integrals 
of motion for the time-like hierarchy, in analogy to the space-like case (see e.g. \cite{FT}). Consequently, we make note here of the first few integrals 
of motion (see also \cite{ACDK}):
\begin{equation}
\begin{aligned}
	H^{(1)} &= \int_{-\tau}^{\tau} \big( u \pi - \hat \pi \hat u \big) dt, \\
	H^{(2)} &= \int_{-\tau}^{\tau} \big( (u\hat u)^2 - u_t \hat{u} - \hat \pi \pi \big) dt, \\
	H^{(3)} &= \int_{-\tau}^{\tau} \big( \hat \pi_{t} \hat u - u_t {\pi} \big) dt, \\
	H^{(4)} &= \int_{-\tau}^{\tau} \big( u_{tt} \hat u + \hat \pi_{t} \hat{\pi} - u\hat u (2 u_t \hat u+ u \hat{u}_t -\hat \pi^2 \hat {u}^2 -
 u^2 {\pi}^2 + 2\hat \pi {\pi} u\hat u) \big) dt,
\end{aligned} \label{Hams}
\end{equation}
where we use the shorthand notation: $F_t = \partial_t F,\ F_{tt} = \partial_{t}^2F$, and so on.
It is also worth noting that this description is in analogy to the relativistic case e.g. sine-Gordon model \cite{FT}, 
where the Hamiltonian is expressed in terms of the sine-Gordon field  $\phi$ and its conjugate $\pi$, where $\pi$ is in turn expressed as 
a time derivative of the field via the corresponding equations of motion.

In addition to the derivation of the time-like charges in involution above we can also compute the corresponding $U$-operators of the time-like 
hierarchy via the expansion in  powers 
of ${1\over \lambda}$  of (\ref{eq:UGen2}).  The pair $ (U^{(k)}, V) $ gives rise to the same equations of motion as Hamilton's equations with 
the Hamiltonian $ H^{(k)} $ associated to the $x_k$ flow. 
We provide below the first few members of the series expansion of ${\mathbb U}$ corresponding to the charges (\ref{Hams})
\begin{equation}
\begin{aligned}
	U^{(1)} &= \lb \begin{matrix}
		1 & 0 \\
		0 & 0
	\end{matrix} \rb, \\
	U^{(2)} &= \lb \begin{matrix}
		\lambda & \hat u \\
		u & 0
	\end{matrix} \rb, \\
	U^{(3)} &= \lb \begin{matrix}
		\lambda^2 - u \hat u & \lambda \hat u+ \pi \\
		\lambda u- \hat \pi & u \hat u
	\end{matrix} \rb, \\
	U^{(4)} &= \lb \begin{matrix}
		\lambda^3 - \lambda u \hat u + \hat \pi \hat u - u \pi & \lambda^2 \hat u + \lambda \pi+ \hat u_t \\
		\lambda^2 u - \lambda \hat \pi - u_t & \lambda u\hat u
 - \hat \pi \hat u + u \pi
	\end{matrix} \rb.
\end{aligned} \label{eq:UMats}
\end{equation}
Having identified both the charges in involution as well as the various $U$-operators, let us focus on the second member of the hierarchy. In particular, let us obtain
via the Hamiltonian $H^{(2)}$ (and the time-like Poisson relations)  and/or the Lax pair $(V,\ U^{(2)})$ the corresponding equations of motion. 
First we obtain (see also \cite{ACDK})
\begin{equation}
\pi(x, t) = \partial_x \hat u(x,t ), ~~~~\hat \pi(x, t) = \partial_x u(x, t),
\end{equation}
and then the equations of motion read as
\begin{equation}
\partial_t u+\partial_x^2 u -2\hat u u^2=0.
\end{equation}
Similarly, for $\hat u$  (but $t \to - t$). This concludes our brief review of the results for dual integrable systems with periodic boundary conditions.

\subsection{Open Boundary Conditions}
\label{sec:OpenBCs}
\noindent We come now to the more interesting scenario where integrable boundary conditions are implemented along the time axis. 
Space-like boundary conditions for NLS and its generalizations  have been investigated (see e.g \cite{Sklyanin, DoFioRa}), 
so we only concern ourselves here with time-like boundaries. This is indeed the first time that such conditions are systematically 
implemented and studied in the context of integrable models. Based on the fundamental relations ($t$-Poisson) we extend the idea of 
Sklyanin's modified monodromy along the time axis. Then via the time-like reflection algebra we are able to construct the generating function of 
time-like Hamiltonians as well as the generating 
function of the $U$-operators in the presence of boundaries. 

The key object in our analysis is Sklyanin's modified monodromy matrix along the time axis, which is given as
\begin{equation}
	\mathcal{T}(\lambda) = T(\lambda) K^{-}(\lambda) \hat T(\lambda), \label{eq:openMonom}
\end{equation}
where $T$ is the time-like monodromy (\ref{tmono}), $\hat T(\lambda) = VT^t(-\lambda) V$ with $V = \mbox{antidiag}(i, -i)$.
Let $ K^{\pm} $ be $c$-number solutions of the classical reflection equation \cite{Sklyanin, Cherednik}:
\begin{eqnarray}
	\Big \{ K^{\pm}_1(\lambda),\ K^{\pm}_2(\mu)\Big \} &=& \Big [ r_{12}(\lambda - \mu),\ K^{\pm}_1(\lambda) K^{\pm}_2(\mu) \Big  ]\nonumber\\
&+& K^{\pm}_1(\lambda) r_{12}(\lambda + \mu) K^{\pm}_2(\mu) - 
K^{\pm}_2(\mu) r_{12}(\lambda + \mu) K^{\pm}_1(\lambda), \nonumber\\ && \label{eq:ReflectEq}
\end{eqnarray}
and consequently ${\cal T}$ is also a solution of the reflection equation. Note that a $c$-number solution of the reflection equation  is a ``non-dynamical'' solution:
$ \Big \{ K^{\pm}_1(\lambda),\ K^{\pm}_2(\mu)\Big \} =0$.
For $ r $ being the Yangian (\ref{YangS}), the most general $ K^{\pm} $-matrices (up to some overall multiplicative factor) are given by \cite{dV-GR}
\begin{equation}
	K^{\pm}(\lambda) = \lb \begin{matrix}
		\lambda + i\xi^{\pm } &i \kappa^{\pm} \lambda  \\
	i\kappa^{\pm} \lambda 	& -\lambda + i \xi^{\pm}
	\end{matrix} \rb, \label{eq:KMat}
\end{equation}
\noindent where $ \xi^{\pm} $, $ \kappa^{\pm} $  are some arbitrary constants\footnote{We could allow these to actually be functions of the evolution variable,
i.e. ``dynamical"  boundary conditions. Doing so would have no effect on our derivations (except to make the expressions 
bulkier by writing in the parameter dependence),
so we choose to ignore this case for now.}.

As in the periodic case we define the generating function of the time-like Hamiltonians for the model with open boundary conditions:
\begin{equation}
\mathcal{G}(\lambda)= \mbox{ln}\big  (\mathfrak{t}(\lambda)\big ), ~~~~ \mathfrak{t}(\lambda)= tr\Big(K^+(\lambda) T (\lambda )K^-(\lambda) \hat T (\lambda)\Big).
\end{equation}
Taking into account the standard decomposition of the monodromy (\ref{deco}), as well the behavior of the ${\mathrm Z}$ matrix as $\lambda \to \infty$
we conclude
\begin{equation}
	\mathcal{G}(\lambda)= {\mathrm Z}_{11} + \hat{{\mathrm Z}}_{11} + \ln{\mathbb{W}_{+}} + \ln{\mathbb{W}_{-}}, \label{eq:OpenHamGen}
\end{equation}
where we define in general $\hat f(\lambda) = f(-\lambda)$, and after taking into account the standard decomposition of the monodromy matrix (\ref{deco}):
\begin{eqnarray}
&&{\mathbb W}^+=  \Big ( \big (1 +\hat {\mathrm W}^t(\tau,  \lambda)\big )V K^+( \lambda)\big (1+{\mathrm W}(\tau, \lambda)\big )\Big )_{11},\\
&&{\mathbb W}^- =  \Big ( \big (1 + {\mathrm W}(-\tau,  \lambda )  \big )^{-1} K^-(\lambda) V\big ( \big (1+\hat {\mathrm W}(-\tau, \lambda ) \big )^{-1} \big )^t \Big )_{11}.
\end{eqnarray}
We focus here on the ``dual  point'' for NLS, that  is when both time and space-like descriptions lead to the same integrable PDEs; this precisely corresponds to the NLS model \cite{ACDK}. 
We first derive the boundary Hamiltonian, expressed in three distinct parts: the bulk Hamiltonian generated by $ ({\mathrm Z}_{11} +\hat{{\mathrm Z}}_{11}) $, 
and the two boundary Hamiltonians $ H_{\pm}^{(2)} $, generated by $ \ln{\mathbb{W}_{\pm}}$ (we have multiplied (\ref{eq:OpenHamGen}) by ${1\over 2}$)
\begin{equation}
{\cal H}^{(2)} = \int_{-\tau}^{\tau} \big( (u \hat u)^2 - u_t \hat u - \hat \pi {\pi} \big) d t + H_+^{(2)}+ H_-^{(2)},
\end{equation}
where the boundary contributions evaluated at $ t= \pm \tau $ are given by
\begin{equation}
H_{+}^{(2)} = \Big ( {\xi^{+} u\over \kappa^+} -{ i \hat \pi \over  \kappa^+} +{u^2\over 2}\Big ) \Big |_{t =\tau}, ~~~~~
H_{-}^{(2)} =   \Big ({\xi^{-}\hat  u\over \kappa^-} -{ i \pi \over \kappa^-} +{\hat u^2\over 2}\Big )\Big |_{t =-\tau}. \label{HB}
\end{equation}

As in the periodic case using the fact that $T$ and $\hat T$ satisfy a quadratic algebra as well as the zero curvature condition 
we can derive in analogy to \cite{AvanDoikou_boundary} the explicit form of the generating function of the $U$-operators:
\begin{eqnarray}
{\mathbb U}_2(t,\lambda) &= &{\mathfrak t}^{-1}(\lambda) tr_1\Big(K_1^+(\lambda) T_1(\tau, t, \lambda) r_{12}(\lambda -\mu)T_1(t, -\tau, \lambda)
 K_1^-(\lambda) \hat T_1(\tau, -\tau, \lambda) \Big )\nonumber \\
&+&{\mathfrak t}^{-1}(\lambda) tr_1\Big (K_1^+(\lambda) T_1(\tau, -\tau , \lambda)K_1^-(\lambda) \hat T_1(t, -\tau, \lambda)  
r_{12}(\lambda +\mu)\hat T_1(\tau, t, \lambda) \Big ),  \nonumber\\
&& t\neq \pm  \tau, \label{bulk1} 
\end{eqnarray}
and at  the boundary points $\pm \tau$:
\begin{eqnarray}
{\mathbb U}_2(\tau, \lambda, \mu) &= &{\mathfrak t}^{-1}(\lambda) tr_1\Big (K_1^+(\lambda) r_{12}(\lambda -\mu)T_1(\tau, -\tau, \lambda) K_1^-(\lambda) 
\hat T_1(\tau, -\tau, \lambda) \Big )\nonumber \\
&+&{\mathfrak t}^{-1}(\lambda) tr_1\Big (K_1^+(\lambda) T_1(\tau, -\tau, \lambda )K_1^-(\lambda) \hat T_1(\tau, -\tau, \lambda) r_{12}(\lambda +\mu)\ \Big )\nonumber \\
& &\label{bou1}\\
{\mathbb U}_2(-\tau, \lambda, \mu) &= &{\mathfrak t}^{-1}(\lambda) tr_1\Big (K_1^+(\lambda) T_1(\tau, -\tau, \lambda) r_{12}(\lambda -\mu) K_1^-(\lambda) 
\hat T_1(\tau, -\tau, \lambda) \Big )\nonumber \\
&+&{\mathfrak t}^{-1}(\lambda) tr_1\Big (K_1^+(\lambda) T_1(\tau, -\tau, \lambda )K_1^-(\lambda) r_{12}(\lambda +\mu) \hat T_1(\tau, -\tau, \lambda)\ \Big ). \nonumber \\
& &\label{bou2}
\end{eqnarray}

Next, we supply the $ U $-matrices associated to the boundary Hamiltonian ${\cal H}^{(2)}$. From expression (\ref{bulk1}) we obtain the familiar bulk NLS $U$-operator
 (we have multiplied expressions (\ref{bulk1})--(\ref{bou2}) by ${1\over 2}$)
\begin{equation}
	U^{(2)} = \begin{pmatrix}
		{\lambda \over 2} & \hat u \\
	u  & -{\lambda \over 2}
	\end{pmatrix} . \label{eq:tUMats_Open(b)}
\end{equation}
\noindent We now turn to the boundary $ U $-matrices, evaluated at $\pm\tau $ from expressions (\ref{bou1}), (\ref{bou2}):
\begin{equation}
	U_{+}^{(2)}= \begin{pmatrix}
		{\lambda\over2 } -{i u \over \kappa^+} & {i\lambda \over \kappa^+} + u + {\xi^+ \over \kappa^+}\\
		u & -{\lambda \over 2} +{iu \over \kappa^+}
	\end{pmatrix}, ~~~~U_{-}^{(2)}=  \begin{pmatrix}
		{\lambda\over2 } -{i \hat u \over \kappa^-} &  \hat u \\
		 {i\lambda \over \kappa^-} + \hat u +{\xi^- \over \kappa^-}& -{\lambda \over 2} +{i \hat u \over \kappa^-}
	\end{pmatrix}.  \label{Open}
\end{equation}
Having at our disposal both the bulk $U$ and boundary $U_{\pm}$-operators we require analyticity conditions at the boundary point following the argument 
for space-like boundary conditions in \cite{AvanDoikou_boundary} 
(see also a relevant discussion in the space picture in \cite{Corrigan}). More precisely,  let
$U_{\pm} = U +\delta U_{\pm}$, by requiring $\delta U_{\pm} =0$ we directly identify the boundary conditions, 
which in this case read as
\begin{eqnarray}
&&\hat u( \tau) ={\xi^+ \over \kappa^+}, ~~~~ u(\tau) =0,  \\
&& \hat u(- \tau) =0, ~~~~ u(-\tau) ={\xi^- \over \kappa^-},
\end{eqnarray}
subject to the extra constraint $\kappa^{\pm},\ \xi^{\pm} \gg 1$, so that the $\lambda$-dependence in the anti-diagonal terms in (\ref{Open})  becomes negligible.

\section{Time-like dressing for the matrix NLS model}
\noindent 
It will be instructive in the frame of the space-time duality picture to describe the ``time-like dressing'' procedure. The idea is the same as the usual dressing scheme,
but now the input is the form of the  $V$-operator together with the general form of the bare operators $U_0^{(n)}$ associated to each $x_n$ flow. 
Having this information at our disposal we are able, as will be transparent in what follows, to identify the hierarchy of $U$-operators.
This picture is admittedly more general compared to the one described so far based on the existence of a classical $r$-matrix, and offers a systematic means to 
produce the time-like hierarchies in the absence of an $r$-matrix or when the associated algebra is too complicated to be practically exploited. 

To illustrate the generality of the process we focus on the matrix NLS model, for which we do not assume any Poisson structure. 
We produce the hierarchy of conserved quantities and Lax pairs based exclusively on the auxiliary linear problem and the  dressing 
transform, and for now we restrict our attention to periodic or vanishing boundary conditions.

\subsection{Non-commutative time Riccati equations}
\noindent Before we proceed with the dressing process and the identification of the hierarchy of Lax pairs we first derive the 
Riccati equations associated to the solution of the auxiliary linear problem and for the general Darboux-dressing transform. 
These as expected  turn out to be equivalent.
Solving the Riccati equation allows the derivation of the conserved quantities corresponding to the integrable hierarchy of interest.

\subsubsection*{Matrix Riccati equations $\&$ conserved quantities}
\noindent Our starting point,  as already mentioned,  is the $V$-operator and the form of the bare $U$ -operators. 
For the matrix NLS model in particular the $V$-operator is  a generalization of  (\ref{NLSb}) 
\begin{equation}
 V(\lambda) = \lb \begin{matrix}
		\frac{ \lambda^2}{2}{\mathbb I}_{N\times N}  -  \hat  u u& \lambda \hat u + \pi\\
		\lambda  u - \hat \pi & - \frac{ \lambda^2}{2}{\mathbb I}_{M\times M}+  u\hat  u
	\end{matrix} \rb. \label{NLSb2}
\end{equation}
where $\hat u,\ \pi$ are $N\times M$ matrices and $u,\hat \pi$ are $M\times N$,
and the bare operators $U_0^{(n)}$ are given as
\begin{equation}
U_0^{(n)} = {\lambda^{n-1} \over 2}\Sigma, ~~~~n \in \{1,2,\ldots \},
\end{equation}
where $\Sigma = \mbox{diag}({\mathbb I}_{N\times N},\  -{\mathbb I}_{M\times M})$. 
Let us  now focus on the time part of the auxiliary linear problem (\ref{ALP}), (\ref{NLSb}) expressed in the block form
\begin{equation}
\partial_t  \lb \begin{matrix}
		\Psi_1\\
		\Psi_2
	\end{matrix} \rb =  \lb \begin{matrix}
		\frac{ \lambda^2}{2}{\mathbb I}_{N\times N}  -  \hat  u u& \lambda \hat u + \pi\\
		\lambda  u - \hat \pi & - \frac{ \lambda^2}{2}{\mathbb I}_{M\times M}+  u\hat  u
	\end{matrix} \rb \lb \begin{matrix}
		\Psi_1\\
		\Psi_2
	\end{matrix} \rb. \label{generalc}
\end{equation}
Let us also define $\Gamma = \Psi_2 \Psi_1^{-1}$, then via (\ref{generalc}) we conclude that $\Gamma$ satisfies the following matrix 
Riccati equation\footnote{Similarly, we could have defined $\hat \Gamma = \Psi_1 \Psi_2^{-1}$, and obtain
\begin{equation}
\partial_t \hat  \Gamma = \lambda \hat u + \pi +(- \hat u u+ {\lambda^2\over 2})  \hat \Gamma +  \hat \Gamma (-u \hat u + {\lambda^2\over 2}) +  
\hat \Gamma(\hat \pi - \lambda u) \hat \Gamma. \label{gg2}
\end{equation}}
\begin{equation}
\partial_t \Gamma = \lambda u -\hat \pi +( u \hat u - {\lambda^2\over 2})  \Gamma + \Gamma (\hat u u -{\lambda^2 \over 2}) -
\Gamma(\pi +\lambda \hat u) \Gamma. \label{gg1}
\end{equation}
Our aim is to solve the Riccati equation (\ref{gg1}). To achieve this we consider the power series expansion $\Gamma = \sum_{k} {\Gamma^{(k)} \over \lambda^k}$, and then 
compute the first few orders of this expansion:
\begin{equation}
\Gamma^{(1)} = u , ~~~~~~\Gamma^{(2)} = -\hat \pi, ~~~~~~\Gamma^{(3)} = -\partial_t u +u\hat u u,  
~~~~~~\Gamma^{(4)} = \partial_t \hat \pi - u\pi  u,\  \ldots \label{gamma}
\end{equation}

We are now in a position to identify the conserved quantities of the model at hand i.e. the multi component generalization of (\ref{Hams}). 
To achieve this let us also take into consideration the $x$-part of the auxiliary linear problem (see also e.g. \cite{DFS1} on similar 
arguments regarding the space-like matrix NLS model). Indeed, the linear equation associated to the $x_n$ flow reads in general as
\begin{equation}
\partial_{x_n}  \lb \begin{matrix}
		\Psi_1\\
		\Psi_2
	\end{matrix} \rb =  \lb \begin{matrix}
		\alpha_n & \beta_n\\
		\gamma_n & \delta_n
	\end{matrix} \rb \lb \begin{matrix}
		\Psi_1\\
		\Psi_2
	\end{matrix} \rb. \label{generald}
\end{equation}
We cross-differentiate (\ref{generalc}), (\ref{generald}) and focus on the first element $\Psi_1$, which leads to
\begin{equation}
\partial_{x_n} \Big (-\hat u u + (\lambda \hat u + \pi)\Gamma \Big ) = \partial_{t} \Big (\alpha_n + \beta_n\Gamma \Big ) 
+ \Big [\alpha_n + \beta_n\Gamma,\  -\hat u u + (\lambda \hat u + \pi)\Gamma \Big ].
\end{equation}
By taking the trace of the expression above and considering the ${1\over \lambda}$ power series expansion 
$\Gamma = \sum_{k} {\Gamma^{(k)} \over \lambda^k}$  we obtain the time-like conserved quantities 
(conserved with respect to $x_n$, and recall we assume periodic or vanishing time boundary conditions):
\begin{equation}
{\cal I}^{(k)} = \sum_{i, j}\int_{-\tau}^{\tau}  \Big ( \hat u_{ij}\Gamma_{ji}^{(k+1)} + \pi_{ij} \Gamma_{ji}^{(k)}\Big )\ dt.
\end{equation}
By substituting the $\Gamma^{(k)}$ from (\ref{gamma}) to the latter expression we identify the first few conserved charges:
\begin{eqnarray}
&& {\cal I}^{(1)} =  \sum_{i, j}\int_{-\tau}^{\tau}  \Big (\pi_{ij} u_{ji} - \hat u_{ij} \hat \pi _{ji} \Big ) dt, \nonumber\\
&& {\cal I}^{(2)} = \sum_{i, j}\int_{-\tau}^{\tau}  \Big ( -\hat u_{ij} \partial_t u_{ji} - \pi_{ij}\hat \pi_{ji} +
\sum_{m, n}\hat u_{im} u_{mn} \hat u_{nj} u _{ji}  \Big ) dt,  \nonumber\\
&& {\cal I}^{(3)} = \sum_{i, j}\int_{-\tau}^{\tau}  \Big (  \hat u_{ij} \partial_t \hat \pi_{ji} - \pi_{ij} \partial_t u_{ji} \Big ) dt, \nonumber\\
&& \ldots
\end{eqnarray}
which are  multi-component generalizations of (\ref{Hams}).

\subsubsection*{Matrix Riccati equations for the dressing transform}
\noindent We now derive the Riccati equations for the general dressing transform and show their equivalence to the equations 
derived from the solution of the time part of the auxiliary linear problem. 
Consider the dressing transform ${\mathrm G}$: 
\begin{equation}
\Psi(\lambda,x,t) = {\mathrm G}(\lambda,x,t) \Psi_0(\lambda) \label{dress1}
\end{equation}
where $\Psi$ satisfies the auxiliary linear problem with Lax pair $U^{(n)},\ V$
and $\Psi_0$ is the bare auxiliary function with the corresponding bare Lax pair $U_0^{(n)},\ V_0$.
In general, via (\ref{dress1}) and the auxiliary linear problem we obtain the fundamental relations for the dressing
\begin{eqnarray}
&&\partial_t{\mathrm G} = V {\mathrm G} - {\mathrm G}V_0  \nonumber\\
&& \partial_{x_n}{\mathrm G} = U^{(n)} {\mathrm G} - {\mathrm G}U_0^{(n)}. \label{fundt}
\end{eqnarray}
We consider the general dressing transform expressed as a formal series expansion
\begin{equation}
{\mathrm G}(\lambda, x,t) = \sum_{k=0}^{\infty}{ {\mathrm g}_k \over \lambda^k} =  \lb \begin{matrix}
		{\mathrm A}_{N\times N}&  {\mathrm B}_{N\times M}  \\
		 {\mathrm C}_{M\times N} & {\mathrm D}_{M\times M}
	\end{matrix} \rb,
\end{equation}
where ${\mathrm g}_k$ are ${\cal N} \times {\cal N}$ (${\cal N} = N+M$) matrices. In particular, ${\mathrm g}_0$ is a 
constant matrix that commutes with $\Sigma$,  and without loss of generality we consider it it to be the identity.

Let us now focus on the $t$-part of the equations above (\ref{fundt}).  Recalling $V$ from (\ref{NLSb}) we conclude
\begin{equation}
\partial_t {\mathrm G} = {\lambda^2 \over 2} \big [\Sigma,\ {\mathrm G} \big] + \lambda {\mathrm X} {\mathrm G} + {\mathrm Y} {\mathrm G},
\label{matrixeq}
\end{equation}
where we define
\begin{equation}
{\mathrm X}=  \lb \begin{matrix}
		0 & \hat u \\
		 u  & 0
	\end{matrix} \rb, ~~~~~{\mathrm Y} =  \lb \begin{matrix}
		- \hat u u& \pi\\
		 -\hat \pi  &  u \hat u
	\end{matrix} \rb.
\end{equation}
We restrict our attention to the first column of the matrix equation (\ref{matrixeq}) and we obtain for the elements  ${\mathrm A},\ {\mathrm C}$
\begin{eqnarray}
&& \partial_t {\mathrm A} = \lambda \hat u {\mathrm C}  - \hat u u {\mathrm A} + \pi {\mathrm C}  \nonumber\\
&&   \partial_t {\mathrm C} = -\lambda^2 {\mathrm C} + \lambda u {\mathrm A} -\hat \pi {\mathrm A} +u\hat u {\mathrm C }. \label{auxilb}
\end{eqnarray}
We define $\Gamma = {\mathrm C} {\mathrm A}^{-1}$, and conclude from (\ref{auxilb}) that it satisfies the time Riccati equation (\ref{gg1}).
Similarly, we could have focused on the second column in the matrix equation (\ref{matrixeq}), and obtained equations analogous 
to (\ref{auxilb}) for ${\mathrm B},\ {\mathrm D}$:
\begin{eqnarray}
&& \partial_t{\mathrm B} = \lambda^2 {\mathrm B} + \lambda \hat u {\mathrm D} - \hat u u {\mathrm B} + \pi {\mathrm D}\nonumber\\
&& \partial_t {\mathrm D} = \lambda u{\mathrm B} - \hat\pi {\mathrm B} + u \hat u {\mathrm D}. \label{BD}
\end{eqnarray}  
We also define 
$\hat \Gamma = {\mathrm B} {\mathrm D}^{-1}$, which in turn satisfies (\ref{gg2}).
It is thus clear that the solution of the $t$-part of the auxiliary linear problem and the general Darboux 
transform lead to the same non-commutative Riccati equations (\ref{gg1}), (\ref{gg2}).

\subsection{Dressing}
\noindent We come now to the derivation of the tower of $U^{(n)}$ operators based on the dressing process, and without {\it a priori}
assuming any algebraic structure for the fields. 
We choose to consider the fundamental dressing transform
\begin{equation}
{\mathrm G} = \lambda {\mathbb I} +  {\mathrm K}.
\end{equation}
This is the simplest case, but nevertheless it provides the whole hierarchy in an efficient way as will be clear in what follows.  
We consider first the $t$-part of relations (\ref{fundt}), which yields
\begin{equation}
\partial_t {\mathrm K} = {\lambda^2 \over 2} \big [\Sigma,\ {\mathrm K}\big ] + \lambda^2 {\mathrm X} + 
\lambda ({\mathrm X} {\mathrm K} +{\mathrm Y}) + {\mathrm Y}{\mathrm K}.
\end{equation}
The latter equation leads to the following set of constraints
\begin{eqnarray}
&& {\mathrm X} = {1\over 2} \big [{\mathrm K},\ \Sigma \big ], \nonumber\\
&&  {\mathrm Y} = -{\mathrm X} {\mathrm K},   \nonumber\\
&& \partial_t {\mathrm K} = {\mathrm Y} {\mathrm K},
\end{eqnarray}
which allow the derivation of the ${\cal N} \times {\cal N}$  matrix ${\mathrm K}$ (see also \cite{DFS1} for a detailed discussion in the space-like picture, 
regarding also the derivation of specific solutions of the fields $u, \hat u$).

Let us now move on with the derivation of $U^{(n)}$, formally expressed as 
\begin{equation}
U^{(n)}(\lambda, x_n,t) = {\lambda^{n-1} \over 2}\Sigma + \sum_{k=0}^{n-2} \lambda^k {\mathfrak w}^{(n)}_k(x_n, t). \label{expansionU}
\end{equation}
From the $x$-part of (\ref{matrixeq}) we conclude
\begin{equation}
\partial_{x_n} {\mathrm K} = {\lambda^{n-1} \over2}\big [\Sigma,\ {\mathrm K}\big ] +\sum_{k=1}^{n-1}\lambda^k {\mathfrak w}^{(n)}_{k-1} + 
\sum_{k=0}^{n-2}\lambda^k {\mathfrak w}^{(n)}_{k}{\mathrm K}.
\end{equation}
We then gather recursion relations for ${\mathfrak w}_k^{(n)}$:
\begin{eqnarray}
&& {\mathfrak w}^{(n)}_{n-2}  ={1\over 2}\big [ {\mathrm K},\ \Sigma \big ]\nonumber\\
&& {\mathfrak w}^{(n)}_{k-1} = -{\mathfrak w}_k^{(n)} {\mathrm K}, ~~~~~k\in \{1, \ldots, n-2 \}\nonumber\\
&& \partial_{x_n} {\mathrm K} =  {\mathfrak w}^{(n)}_0 {\mathrm K}.
\end{eqnarray}

Solving the recursion relations yields precisely the members of the hierarchy.
In particular, for $n=1$ we trivially find $U^{(1)} = {\Sigma \over 2}$, and for the first couple of terms we deduce:\\
\begin{itemize}
\item $n=2$
\begin{equation}
{\mathfrak w}_{0}^{(2)} =   \lb \begin{matrix}
		0& \hat u\\
		 u  &  0
	\end{matrix} \rb,\   
\end{equation}
\item $n=3$
\begin{equation}
{\mathfrak w}_1^{(3)}= {\mathfrak w}_{0}^{(2)}, ~~~~~~~{\mathfrak w}_{0}^{(3)} =   \lb \begin{matrix}
		- \hat u u& \pi\\
		 -\hat \pi  &  u \hat u
	\end{matrix} \rb,\ 
\end{equation}
\item $n=4$
\begin{equation}
{\mathfrak w}_{2}^{(4)} = {\mathfrak w}_{0}^{(2)}, ~~~~~~{\mathfrak w}_{1}^{(4)} = {\mathfrak w}_{0}^{(3)} ,
~~~~~~~{\mathfrak w}_{0}^{(4)} =   \lb \begin{matrix}
		 \hat u \hat \pi -  \pi u&  \partial_t \hat u \\
		 - \partial_t  u & - \hat \pi \hat u + u \pi
	\end{matrix} \rb.
\end{equation}
\end{itemize}
Given the expressions above and the expansion (\ref{expansionU}) we obtain
\begin{equation}
\begin{aligned}
	U^{(1)} &= {1\over 2} \Sigma, \\
	U^{(2)} &= \lb \begin{matrix}
		{\lambda \over 2}{\mathbb I}_{N\times N} & \hat u \\
		u & -{\lambda\over 2}{\mathbb I}_{M\times M}
	\end{matrix} \rb, \\
	U^{(3)} &= \lb \begin{matrix}
		{\lambda^2\over 2} {\mathbb I}_{N\times N}- \hat u u & \lambda \hat u+ \pi \\
		\lambda u- \hat \pi & -{\lambda^2 \over 2}{\mathbb I}_{M\times M} + u \hat u
	\end{matrix} \rb, \\
	U^{(4)} &= \lb \begin{matrix}
		{\lambda^3\over 2}{\mathbb I}_{N\times N} - \lambda \hat u u + \hat u \hat \pi -  \pi u& \lambda^2 \hat u + \lambda \pi+ \partial_t \hat u \\
		\lambda^2 u - \lambda \hat \pi - \partial_t u &- {\lambda^3\over 2}{\mathbb I}_{M\times M} +\lambda u\hat u
 - \hat \pi \hat u + u \pi
	\end{matrix} \rb, \\
\ldots
\end{aligned} \label{eq:UMats2}
\end{equation}
which are the matrix generalizations of (\ref{eq:UMats}) (recall that for our chosen model 
$\hat u,\  \pi$ are $N \times M$ matrices, and $u,\ \hat \pi$ are $M\times N$ matrices). 
For the Lax pair $(V,\ U ^{(2)})$ in particular we obtain
\begin{equation}
\pi(x, t) = \partial_x \hat u(x,t ), ~~~~\hat \pi(x, t) = \partial_x u(x, t).
\end{equation}
and we recover the matrix NLS equation
\begin{equation}
\partial_t u+\partial_x^2 u -2u\hat u u=0. \label{NLSmatrix}
\end{equation}
Note that expressions (\ref{eq:UMats}) were computed from the universal algebraic formula (\ref{eq:UGen}) 
based on the existence of the classical $r$-matrix. 
Here, however we have assumed no algebraic structure, instead by only implementing the 
dressing transform we were able to obtain the time-like matrix NLS hierarchy (\ref{eq:UMats2}).

\section{Discussion}
\noindent
In the first part of our presentation we have used the strong integrability argument based on the existence of the classical $r$-matrix and the underlying Poisson structure. 
To incorporate integrable time-like boundary conditions we introduced the notion of the reflection equation and we considered representations of the reflection algebra i.e. 
Sklyanin's modified monodromy matrices, along the time axis (\ref{tmono}),  (\ref{eq:openMonom}). From the trace of the time-like modified monodromy matrices we were able 
to derive the boundary time-like conserved quantities. Moreover, we identified the generating function of the bulk and boundary $U$-operators for given boundary
conditions defined by $c$-number solutions of the reflection equation.  The boundary conditions on the fields were also identified by imposing suitable analyticity conditions 
on the boundary $U$-operators.

To illustrate how the time-like hierarchy can be obtained in the absence of an $r$-matrix we derived the time-like matrix NLS hierarchy for periodic boundary conditions, 
based exclusively on the auxiliary linear problem and the dressing transform. We were able to generalize previous results on the derivation of the space-time duality \cite{ACDK}, 
in the absence of  a classical $r$-matrix using the frame of the  ``time-like'' dressing transform.

The derivation of space and time-like boundary conditions based exclusively on the existence of a Lax pair, expressed in terms of differential equations, and the dressing process,  
together with a suitable boundary Gelfand-Levitan-Marchenko equation,  in the spirit of \cite{ZakharovShabat1, ZakharovShabat2, DFS1} is one of the next key issues to address. 
The time-like dressing should involve time-like differential and integral operators as dressing transformations, which should in turn provide the solutions of the associated integrable PDEs. 
We hope to report on these and relevant important open issues mentioned throughout the text  in forthcoming publications.


\subsection*{Acknowledgments}
\noindent 
A.D. wishes to thank LPTM, University of Cergy-Pontoise, where part of this work was completed, and J. Avan for kind hospitality 
and useful comments and suggestions. I.F. is supported by EPSRC via a DTA scholarship. S.S. is supported by Heriot-Watt University via a J. Watt scholarship.

\appendix
\section{Time Riccati equation: ${\mathrm W},\ {\mathrm Z}$ matrices}
\noindent In this Appendix we compute the first few members of the expansion $\sum {{\mathrm W}^{(n)} \over \lambda^n},\ \sum {{\mathrm Z}^{(n)} \over \lambda^n}$ 
by solving the time Riccati equation \eqref{tRiccati}. Indeed, solving the time Riccati equation at each order of the ${1\over \lambda}$ expansion we obtain:
\begin{equation}
\begin{aligned}
	{\mathrm W}^{(1)} &= \lb \begin{matrix}
		0 & -\hat u \\
		u & 0
	\end{matrix} \rb, \\
	{\mathrm W}^{(2)} &= \lb \begin{matrix}
		0 & -\pi \\
		-\hat \pi & 0
	\end{matrix} \rb, \\
	{\mathrm W}^{(3)} &= \lb \begin{matrix}
		0 & -\hat u_t - \hat u u\hat u  \\
		-u_t + u u\hat u & 0
	\end{matrix} \rb, \\
	{\mathrm W}^{(4)} &= \lb \begin{matrix}
		0 & -\pi_t - \hat \pi \hat u^2 \\
		\hat \pi_{t} - u^2 \pi  & 0
	\end{matrix} \rb, \\
	{\mathrm W}^{(5)} &= \lb \begin{matrix}
		0 & u \pi^2 - \hat{u}_{tt} - u_t \hat{u}^2 - 2\hat u \big ( u \hat u_t  - \pi \hat \pi \big ) \\
		u_{tt} -\hat \pi^2 \bar{\psi} - u^2 \hat{u}_t - 2u \big (u_t \hat{u} + \pi\hat {\pi}\big ) & 0
	\end{matrix} \rb.
\end{aligned} \label{eqs:tWExprs}
\end{equation}

\noindent We can use the expressions above to calculate the first few elements in the expansion of ${\mathrm Z}$, via the equation for the diagonal part $Z$ (\ref{riccati1})
\begin{equation}
\begin{aligned}
	{\mathrm Z}^{(-2)} &= \lb \begin{matrix}
		\tau & 0 \\
		0 & -\tau
	\end{matrix} \rb, \\
	{\mathrm Z}^{(1)} &= \lb \begin{matrix}
		\int_{-\tau}^{\tau} \big (u \pi - \hat u \hat \pi) d t & 0 \\
		0 & -\int_{-\tau}^{\tau} \big (u \pi - \hat u \hat \pi \big  ) d t
	\end{matrix} \rb, \\
	{\mathrm Z}^{(2)} &= \lb \begin{matrix}
		\int_{-\tau}^{\tau} \big ((u\hat u )^2- u_t \hat{u} - \pi \hat \pi\big ) d t & 0 \\
		0 & -\int_{-\tau}^{\tau} \big ((u\hat u)^2  + u \hat u_t- \pi \hat \pi \big) d t
	\end{matrix} \rb, \\
	{\mathrm Z}^{(3)} &= \lb \begin{matrix}
		\int_{-\tau}^{\tau}\big (\hat \pi_{t} \hat u - u_t \pi \big) d t & 0 \\
		0 & -\int_{-\tau}^{\tau}\big  (u {\pi}_{t} - \hat \pi \hat {u}_t \big ) d t
	\end{matrix} \rb, \\
	{\mathrm Z}^{(4)} &= \int_{-\tau}^{\tau} \lb \begin{matrix}
		u_{tt} \hat u  + \hat \pi_t \pi  - u\hat u \big (2u_t \hat{u} + u \hat {u}_t\big ) & 0 \\
		0 & -u \hat{u}_{tt} + \hat \pi  \pi_t  - u\hat u \big (2u \hat {u}_t + u_t \hat u \big )
	\end{matrix} \rb d t \\
	&\qquad- \int_{-\tau}^{\tau} \lb \hat \pi^2 \hat {u}^2 + u^2 \pi^2 - 2\pi \hat \pi u\hat u  \rb d t \lb \begin{matrix}
		1 & 0 \\
		0 & -1
	\end{matrix} \rb.
\end{aligned} \label{eqs:tZExprs}
\end{equation}

\end{document}